\documentclass{aa}
\usepackage{psfig}
\usepackage{epsfig}
\def \sax {BeppoSAX}
\def \src {X\thinspace2127+119}
\def \glob {M15}

\def \nh {N${\rm _H}$}
\def \hcm {atom cm$^{-2}$}
\def \ferg {erg cm$^{-2}$ s$^{-1}$}
\def \arcsec {\hbox{$^{\prime\prime}$}}

\def \rchisq {$\chi_{\nu} ^{2}$}
\def\approxgt{\mathrel{\hbox{\rlap{\lower.55ex \hbox {$\sim$}}
        \kern-.3em \raise.4ex \hbox{$>$}}}}
\def\approxlt{\mathrel{\hbox{\rlap{\lower.55ex \hbox {$\sim$}}
        \kern-.3em \raise.4ex \hbox{$<$}}}}

\begin{document}

\thesaurus{(02.01.2; 08.09.2 X2127+119; 08.14.1; 10.07.3 M15 (NGC 7078); 13.25.3)}

\title{BeppoSAX spectroscopy of the NGC\thinspace7078
(M\thinspace15) globular cluster X-ray source \src\ }

\author{L. Sidoli
        \and A.N. Parmar
        \and T. Oosterbroek 
}
\offprints{L. Sidoli (lsidoli@astro.estec.esa.nl)}

\institute{
       Astrophysics Division, Space Science Department of ESA, ESTEC,
       Postbus 299, NL-2200 AG Noordwijk, The Netherlands
}
\date{Received 21 March 2000; Accepted: 20 June 2000}

\maketitle

\markboth{BeppoSAX observation of \src}{BeppoSAX 
observation of \src}

\begin{abstract}

Results of a 1999 November 16--17 BeppoSAX observation 
of the low-mass X-ray binary
\src\ located in the globular cluster \glob\ are presented.
The system is believed to be one where the
central neutron star is normally obscured by the accretion disk,
and only X-rays scattered into our line of sight by an
extended accretion disk corona (ADC) are observed.
The 0.1--10~keV lightcurve is energy dependent and
shows two partial eclipses
separated by the 17.1~hr orbital period.
The 0.1--100~keV spectrum is unusually complex, 
but can be successfully modeled 
using a partially covered power-law and disk-blackbody model. 
Together with a column consistent with the interstellar value to 
\glob, $\sim$60\% of the source is covered by an additional
column of $\sim$$10^{22}$~atom~cm$^{-2}$. The absorbed component 
may be X-rays that pass
through the outer layers of the accretion disk.
The energy dependent intensity variations by a factor of $\sim$2 
may be modeled as due to a changing 
normalization of the disk-blackbody. None of the 
other spectral parameters appear to clearly depend 
on luminosity. The same spectral model is
also able to fit an archival ASCA spectrum.
We demonstrate that during the luminous ($\sim$ Eddington) X-ray
burst observed from \src\ by {\it Ginga}, material
located in the outer regions of the accretion disk could have been
temporarily ionized, so allowing the central neutron star
to be viewed directly.

\keywords{Accretion, accretion disks -- Stars: individual: \src\ 
-- Stars: neutron -- Globular cluster: individual: \glob\ (NGC\thinspace7078)
-- X-rays: general}

\end{abstract}

\section{Introduction}
\label{sect:intro}

The X-ray source \src\ is one of 12 bright 
(${\rm L_x > 10^{36}}$~erg~s$^{-1}$) low-mass 
X-ray binaries (LMXRBs) located
within globular clusters. The X-ray source is located within 2\arcsec\ of 
the core of \glob\ (NGC\thinspace7078) and is coincident with the 
variable blue V $\sim 15$
star AC\thinspace211 (Auri\`ere et al. \cite{a:84}) which exhibits 
prominent optical and UV emission and absorption lines 
(Charles et al. \cite{c:86}; Downes et al. \cite{d:96}).
The optical lightcurve exhibits the largest amplitude variations of any
LMXRB and is modulated with a period of 17.1~hr
(Ilovaisky et al. \cite{i:93}; hereafter I93). 
The X-ray lightcurve is weakly modulated
with the same period, but shows a more complex
behavior (see e.g., Homer \& Charles \cite{h:98};   hereafter HC98).
The current model for the system consists of an X-ray binary
viewed at a high inclination angle such that the central X-ray source
is obscured by an accretion disk and only X-rays scattered into the line
of sight by an extended accretion disk corona (ADC), or wind, are
observed. This obscuration accounts for the low 
${\rm L_x/L_{opt}} \sim 20$ ratio for \src, typical of other ADC
sources. Variations in the disk rim height and obscuration
by the companion produce the observed modulation.
The detection
of a luminous ($4.5 \times 10^{38}$~erg~s$^{-1}$ 
at the peak,
assuming isotropic emission) X-ray burst from
\src\ by {\it Ginga} confirms that the compact object is a neutron star and 
implies that, at times,
the neutron star is viewed directly (Dotani et al. \cite{d:90};
van Paradijs et al. \cite{v:90}).
The mean \glob\ metallicity is only $\sim$0.01 solar 
(Geisler et al. \cite{g:92}; Sneden et al. \cite{s:91})  
although this does not necessarily
mean that the accreted material in the \src\ system 
is strongly metal depleted since the
companion star may have undergone a non-standard evolution during which
its envelope composition was altered.

The X-ray spectrum of \src\ is complex and varies on time scales
shorter than the orbital period.
Hertz \& Grindlay (\cite{hg:83}) combined data from the {\it Einstein}
Monitor Proportional Counter (MPC) and High-Resolution Imager
instruments and found that a two component model was required. 
One component is a thermal bremsstrahlung with a temperature, kT,
of $\sim$5~keV and the other appreciably softer. The source
exhibited a strong anti-correlation between luminosity 
and absorption, ${\rm N_H}$.
The EXOSAT Channel Multiplier Array and Medium Energy Detector Array
spectra of Callanan et al. (\cite{c:87}) 
are well fit by a two component
model consisting of a $\sim$1~keV
blackbody together with a power-law with a photon index, $\alpha$, of
1.6, or a blackbody with a similar kT together with a 7--15 keV thermal 
bremsstrahlung. 
The ${\rm N_H}$ was variable in the range 
0.15--3$\times 10^{21}$~atom~cm$^{-2}$ and showed a correlation with
luminosity, in contrast to the result of Hertz \& Grindlay (\cite{hg:83}).
Christian et al. (\cite{c:97}) present results from
the {\it Einstein} Solid State Spectrometer and MPC instruments
and the broadband X-ray telescope (BBXRT). They find that the
continuum is well described by the same model used for the EXOSAT
data  (a power-law plus a 1 keV blackbody) 
with  some evidence for 
the same anti-correlation as Hertz \& Grindlay (\cite{hg:83}).
When the best-fit values of ${\rm N_H}$ are plotted against
orbital phase, $\Phi$, a factor $\sim$10 increase in ${\rm N_H}$
is evident around $\Phi = 0.4-0.7$ (where $\Phi = 0.0$   
corresponds to the center of the partial eclipse of the
X--ray source by the companion star).
If a spectral model with abundances fixed
at the mean \glob\ value is used to model this extra absorption, then
the fit is significantly poorer, with the model over-estimating
the observed data at $\sim$1.2~keV (see Christian et al. \cite{c:97}, 
Fig.~5). 
No narrow Fe line is seen in the BBXRT data, with a 3$\sigma$ upper 
limit of 150~eV. 

We report here on a BeppoSAX observation of \src\ made as part of a
systematic study of luminous globular cluster X-ray sources
(see Guainazzi et al. \cite{g:00} for an overview).
Results for the sources located in Terzans~1 and 2 and NGC\thinspace6441 
are to be found
in Guainazzi et al. (\cite{g:98}, \cite{g:99}) and Parmar et al.
(\cite{p:99}), respectively. 
\glob\ has a low reddening (${\rm E_{B-V}} = 0.10 \pm{0.01}$; Durrell
\& Harris \cite{dh:93}) which corresponds to  
${\rm N_H}\sim7 \times 10^{20}$~atom~cm$^{-2}$, 
using the relation between
${\rm A_v}$ and ${\rm N_H}$ in Predehl \& Schmitt (\cite{ps:95}).
This low ${\rm N_H}$ means that the source is particularly interesting 
at energies $\approxlt$1~keV. We compare the BeppoSAX
results with an unpublished ASCA spectrum obtained in 1995. 
We note that EUV emission has been detected from \glob\ which
may originate from \src\ (Callanan et al. \cite{c:99}). 

\section{Observations}
\label{sect:obs}

Results from the co-aligned Low-Energy Concentrator Spectrometer (LECS;
0.1--10~keV; Parmar et al. \cite{p:97}), the Medium-Energy Concentrator
Spectrometer (MECS; 1.8--10~keV; Boella et al. \cite{b:97}),
and the Phoswich Detection System (PDS; 15--300~keV; 
Frontera et al. \cite{f:97}) on-board \sax\ are presented.
Due to technical reasons the HPGSPC was not operated.
The MECS consists of two grazing incidence
telescopes with imaging gas scintillation proportional counters in
their focal planes. The LECS uses an identical concentrator system as
the MECS, but utilizes an ultra-thin entrance window and
a driftless configuration to extend the low-energy response to
0.1~keV. 
The non-imaging
PDS consists of four independent units arranged in pairs each having a
separate collimator. Each collimator was alternatively
rocked on- and 210\arcmin\ off-source every 96~s during 
the observation.

The region of sky containing \src\ was observed by \sax\
on 1999 November 16 01:15 to November 17 00:50 UTC.
Good data were selected 
when the instrument
configurations were nominal, using the SAXDAS 2.0.0 data analysis package.
LECS and MECS data were extracted centered on the position of \src\ 
using radii of 8\arcmin\ and 4\arcmin, respectively.
The exposures 
in the LECS, MECS, and PDS instruments are 11.1~ks, 35.5~ks,
and 16.6~ks, respectively. 
Background subtraction for the imaging instruments
was performed using standard files, but is not critical for such a
bright source. 
Background subtraction for the PDS used data
obtained during intervals when the collimators were offset from the 
source. 

The BeppoSAX data is compared with results from the Solid State Imaging
Spectrometers SIS0 and SIS1 (0.6--10~keV), 
on-board ASCA (Tanaka et al. \cite{t:94}). 
ASCA observed \src\ between 1995 May 16 00:54 and May 17 03:50~UTC.
Timing results from this observation are reported in HC98. 
The SIS exposure is 25.0~ks using 1-CCD BRIGHT2 mode with the maximum
available telemetry allocation. 
All data were screened and processed using the
standard Rev2 pipeline. The source count rates of 
$\approxlt$9.0~s$^{-1}$~SIS$^{-1}$ mean that pulse pile-up is unlikely
to be significant.

\section{Analysis and results}
\label{sect:analysis}

\subsection{BeppoSAX lightcurve}
\label{subsect:lc}


\begin{figure*}
\centerline{\psfig{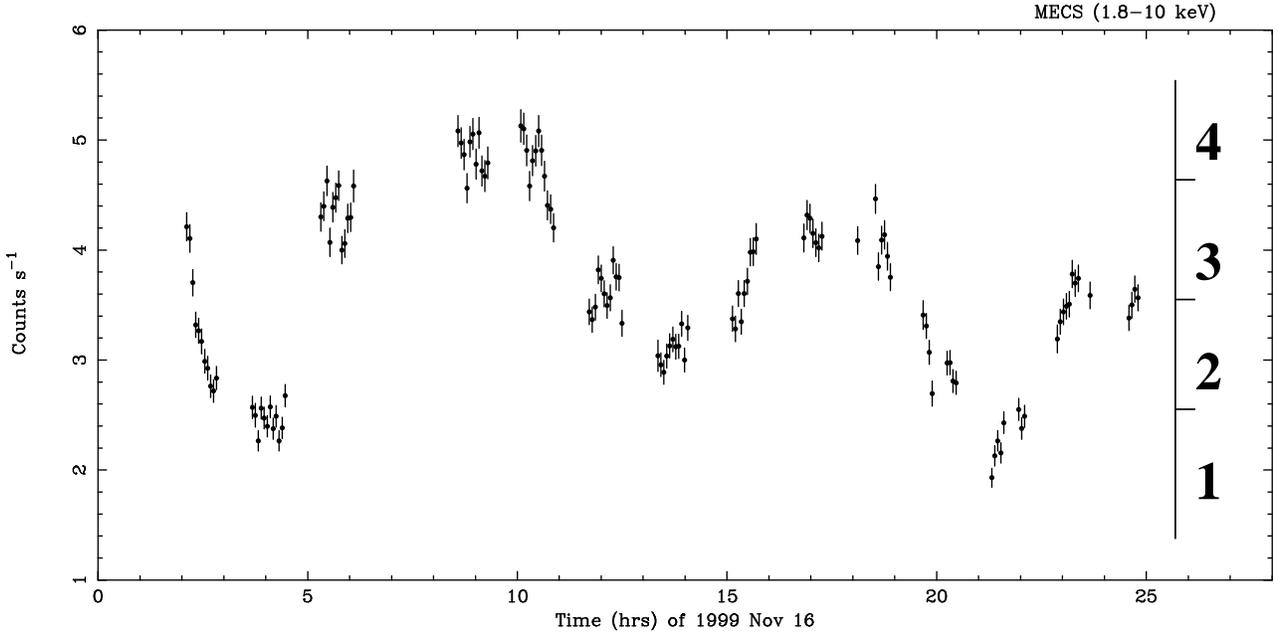}}
     \caption[]{The 1.8--10 keV lightcurve of \src.
     The 4 intensity intervals used in the spectral analysis 
     (see Sect.~\ref{sect:intspe}) 
     are indicated. The bin time is 256 s}
   \label{fig:int}
\end{figure*}


\begin{figure*}
\vskip 0.3truecm
   \centerline{\psfig{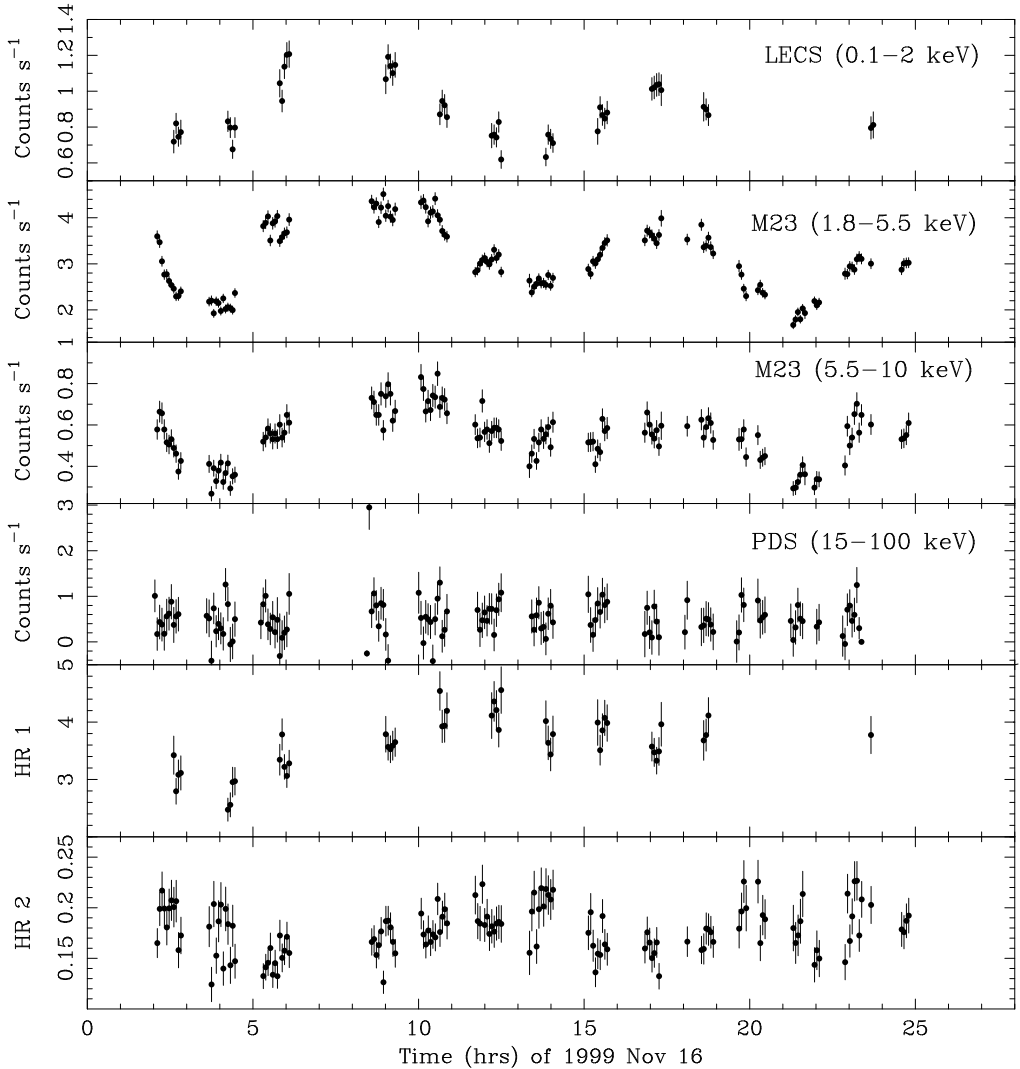}}
\vskip 1truecm
   \caption[]{\src\ lightcurves in different energy bands and
              hardness ratios (lower 2 panels). HR1 is 
MECS 1.8--5.5 keV counts / LECS 0.1--2 keV counts and HR2  
MECS 5.5--10 keV counts / MECS 1.8--5.5 keV counts. 
The bin time is 256~s 
}

   \label{fig:all}
\end{figure*}


\begin{figure*}
\centerline{\psfig{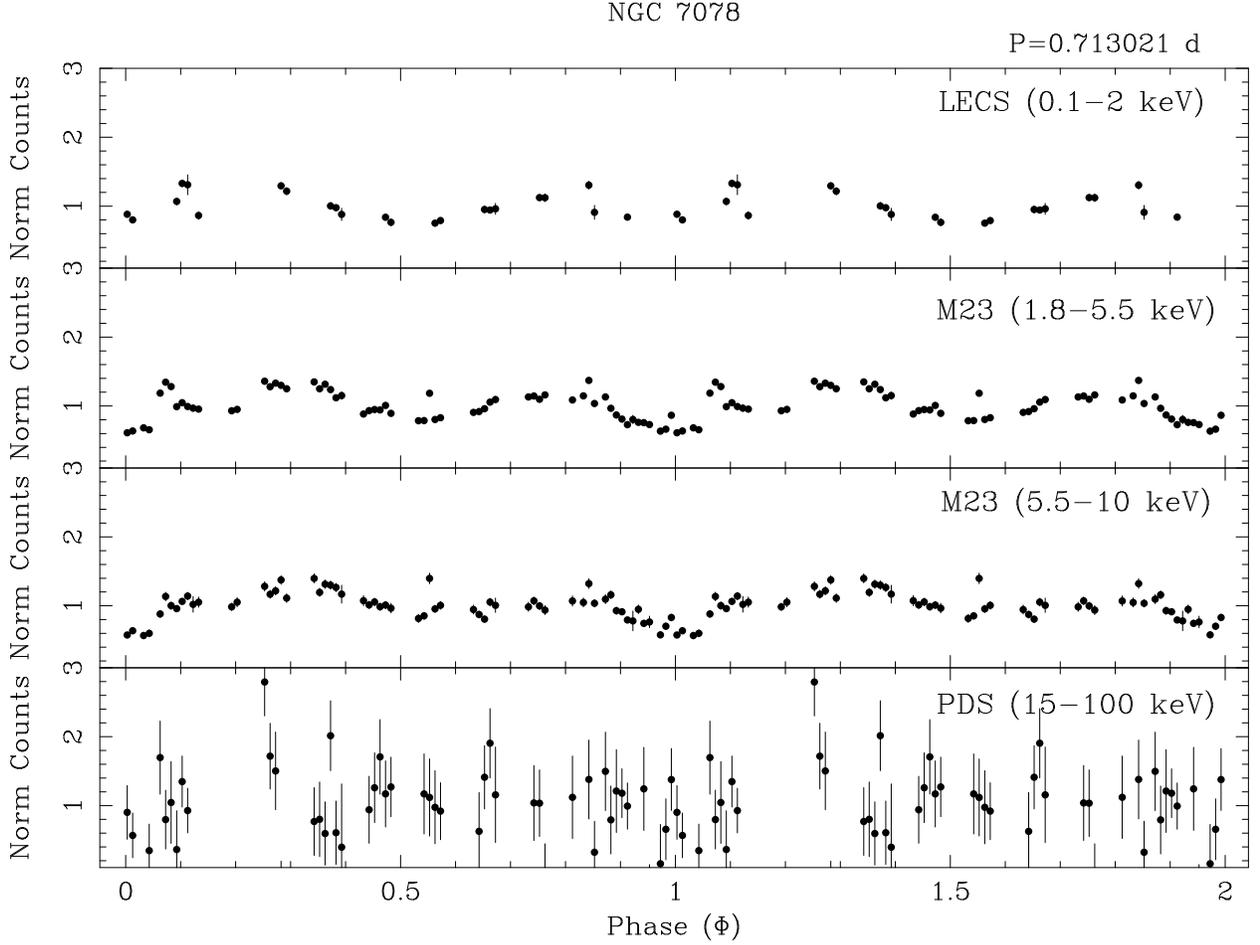}}
    \caption[]{\src\ lightcurves scaled to show the same range of variability}
   \label{fig:folded}
\end{figure*}


The BeppoSAX observation duration of 85~ks includes one entire orbital 
cycle of \src. Fig.~\ref{fig:int} shows the 1.8--10~keV 
background subtracted MECS lightcurve of \src\ with a binning time of 256~s.
Fig.~\ref{fig:all} shows background subtracted lightcurves in the energy 
ranges 
0.1--2~keV (LECS), 1.8--5.5~keV (MECS), 5.5--10~keV (MECS) and 
15--100~keV (PDS) together with 
hardness ratios (1.8--5.5~keV counts divided by 0.1--2~keV counts, 
and 5.5--10~keV counts divided by 1.8--5.5~keV counts).
There is strong variability in the energy range
0.1--10~keV, with two deep minima separated by $\sim$17~hr and a 
  broad 
secondary minimum 
that occured approximately midway between the two deep minima.
The 3$\sigma$ upper limit to any variability in the 15--100~keV
PDS lightcurve (rms$<$0.61) is consistent with the amount of 
variability seen at lower energies.
The complex variations in hardness ratio suggest that it may be important to
perform a spectral analysis in different source intensity ranges 
(see Sect.~\ref{subsect:spectrum}).

\subsection{Eclipse timing}
\label{subsect:eclipse}

Recently HC98 refined the orbital ephemeris of I93
using results from all available X-ray lightcurves from HEAO-1 (1977)
to R-XTE (1998). At the time of the BeppoSAX observation, the
difference in $\Phi$ between the two ephemerides is $\sim$0.05.
HC98 explained this difference either as an
up-dated period of 0.713021~days (7$\sigma$ larger than the
I93 value of 0.713014~days), or by the
introduction of a period derivative of ${\rm \dot{P}_{orb}/P_{orb} =
9\times10^{-7}}~{\rm yr^{-1}}$.
Fig.~\ref{fig:folded} shows lightcurves in
different energy bands plotted against $\Phi$ using the
HC98 linear ephemeris.
There is good agreement
between the BeppoSAX data and the refined ephemeris.
In order to investigate whether the BeppoSAX data can be used
to further refine the ephemeris we performed a detailed
timing analysis.

The occurrence time of the first minimum in the BeppoSAX lightcurve
is 2,451,498.6614$\pm$0.0044 JD$_\odot$,
which corresponds to cycle 5200 using the
I93 epoch. This eclipse time was calculated 
by folding the 1.8--10~keV MECS data at the HC98 linear ephemeris 
and selecting an interval of 0.2 in phase
around the partial eclipse. A fit with a model consisting of 
a constant plus a 
gaussian was made to this data. The uncertainty in the eclipse time was
obtained from $\Delta \chi^{2}$=1.0 after scaling the errors such that 
\rchisq\ was $\sim$1. 
The partial eclipse
arrival time extrapolated to cycle 5200 starting from this same epoch,
but using the new HC98 period of
0.713021~days is 2,451,498.6774$\pm$0.023 (JD$_\odot$), a difference
of 0.016$\pm$0.023~days. Note that, since HC98 and I93 apparently
did not apply a barycentric correction, we have included this
correction (which is however much smaller than the
uncertainties). Using the quadratic ephemeris of HC98 (for which
the difference with respect to the I93 ephemeris is estimated from
HC98 Fig.~2 and amounts to $\sim$0.07 cycle) we obtain an
expected occurrence time of 2,451,498.691$\pm$0.03 JD$_\odot$, a
difference of 0.03$\pm$0.03~days. Similarly, the predicted partial eclipse
center using
the original I93 ephemeris is 2,451,498.6410$\pm$0.023 (JD$_\odot$),
resulting in a difference of 0.020$\pm$0.023~days.

Thus, our measure of the partial eclipse occurrence is in agreement
with the 3 previously published ephemerides and does not allow us to
distinguish between them.

\subsection{BeppoSAX spectrum}
\label{subsect:spectrum}

\subsubsection{Overall spectrum}
\label{subsect:overall}
 
The overall spectrum of \src\ was first investigated by simultaneously
fitting data from the LECS, MECS and PDS 
using {\sc xspec} version 11.01.
All spectra were rebinned using standard procedures.
Data were selected in the energy ranges
0.1--4.0~keV (LECS), 1.8--10~keV (MECS),
and 15--100~keV (PDS) 
where the instrument responses are well determined and sufficient
counts obtained. 
This gives
background-subtracted count rates of 2.1, 3.7, and 1.0~s$^{-1}$ 
for the LECS, MECS, and PDS, respectively.
The photo-electric absorption
cross sections of Morrison \& McCammon (\cite{m:83}) and the
abundances of Anders \& Grevesse (\cite{a:89}) are used throughout.
Factors were included in the spectral fitting to allow for normalization 
uncertainties between the instruments. These factors were constrained
to be within their usual ranges during the fitting. All spectral uncertainties
and upper-limits are given at 90\% confidence.

Initially, simple models were tried, including absorbed power-law,
thermal bremsstrahlung and cutoff power-law 
models, all yielding unacceptable results for the broad band spectrum.
When a power-law model is used, 
examination of the   residuals   shows 
a curved spectrum in the energy range 2--10 keV and a structured 
soft excess below about 1 keV.
Several different combinations of spectral
model were tried next, always including 
a power-law, which is required by the PDS data. Both
a blackbody and a bremsstrahlung were added to the power-law
in order to account for the curvature of the 2--10~keV spectrum,
but always gave unacceptable results, primarily due to 
the structured residuals below 1~keV. 

An absorbed (\nh\ $\sim 10^{21}$ \hcm) 
multicolour disk-blackbody (the {\sc diskbb} 
model in {\sc xspec}) 
(with kT${\rm _{in}}\sim$~1.8~keV) 
together with an absorbed power-law with $\alpha \sim1.7$
gives significantly better results, 
but again low-energy residuals are present.
The multicolour disk-blackbody model of Mitsuda et al. (\cite{m:84}) and 
Makishima et al. (\cite{m:86}) 
assumes that the gravitational
energy released by the accreting material is locally
dissipated into blackbody radiation, that the accretion flow is
continuous throughout the disk and that the effects
of electron scattering on the spectrum are negligible.
There are only two parameters in the model:
R${\rm _{in} (\cos \theta)^{0.5} / d_{10}}$ where
R$_{\rm in}$ is the innermost radius of the disk,
$\theta$ the inclination
angle of the disk, d$_{10}$ the source distance in units of 10~kpc,
and kT$_{\rm in}$ the blackbody effective temperature at R$_{\rm in}$.
The limitations of this model are discussed in 
Merloni et al. (\cite{mfr:99}).

The low metallicity of the hosting globular cluster ($\sim$0.01
solar) prompted us to try using
a photoelectric absorption model with variable abundances ({\sc vphabs} in 
{\sc xspec}), together with a  photoelectric absorption fixed at the
interstellar value of $7 \times 10^{20}$~atom~cm$^{-2}$. 
Different combinations of linked abundances were tried. 
First, fixing the He abundance at the
cosmic value
and linking all the abundances of the metals together as a   single
 parameter.
Then, linking C, N and O together and grouping the other
metals (Ne, Na, Mg, Al, Si, S, Cl, Ar, Ca, Cr, Fe, Co, Ni) as another 
parameter, both fixing the abundance of the metals 
at 0.01 of the solar value and
letting it vary freely. In all cases unacceptable results are obtained.
No K$_{\alpha}$ iron line is present with a 90\% confidence upper limit 
to the equivalent width of a narrow line of 50~eV.

The positive residuals below 1~keV suggest the presence of 
either (1) partial covering, (2) an ionized absorber 
or (3) an additional soft component.
Thus, we first used the same model as before (a power-law plus 
a disk-blackbody), but allowed the absorption of each component
to vary separately, while being constrained to be not less than the
galactic value. Significantly better fits were obtained when the
disk-blackbody suffered extra absorption, and so the absorption
of the power-law component was set equal to the galactic value, 
${\rm N_{gal}}$ ($7 \times 10^{20}$~atom~cm$^{-2}$). 
The model can be written as follows:

\[
{\rm Intensity = e^{-\sigma N_{gal}} \;( I_{DBB}  e^{-\sigma
N_H} \rm +\;  I_{PL}),}
\]
where $\rm {I_{DBB}}$ and $\rm {I_{{\rm PL}}}$ are the normalizations of the
disk-blackbody and power-law components, and
${\rm N_H}$ is the additional column density of the disk-blackbody.
The best-fit values are
\nh\ = $(7.0  ^{+1.00} _{-0.60}) \times 10^{21}$ atom cm$^{-2}$, 
kT$_{{\rm in}}$ = $1.72 ^{+0.04} _{-0.03}$~keV, 
R$_{\rm in} (\cos \theta)^{0.5} / d_{\rm 10~kpc}$ =  
$1.2 \pm 0.05$~km, $\alpha = 1.84 \pm 0.05$.
Although the fit is not formally acceptable with a $\chi ^2$ of 177 for 125
degrees of freedom (dof), the overall shape of the spectrum
is reasonably well accounted for by this model.
 
\begin{figure*}
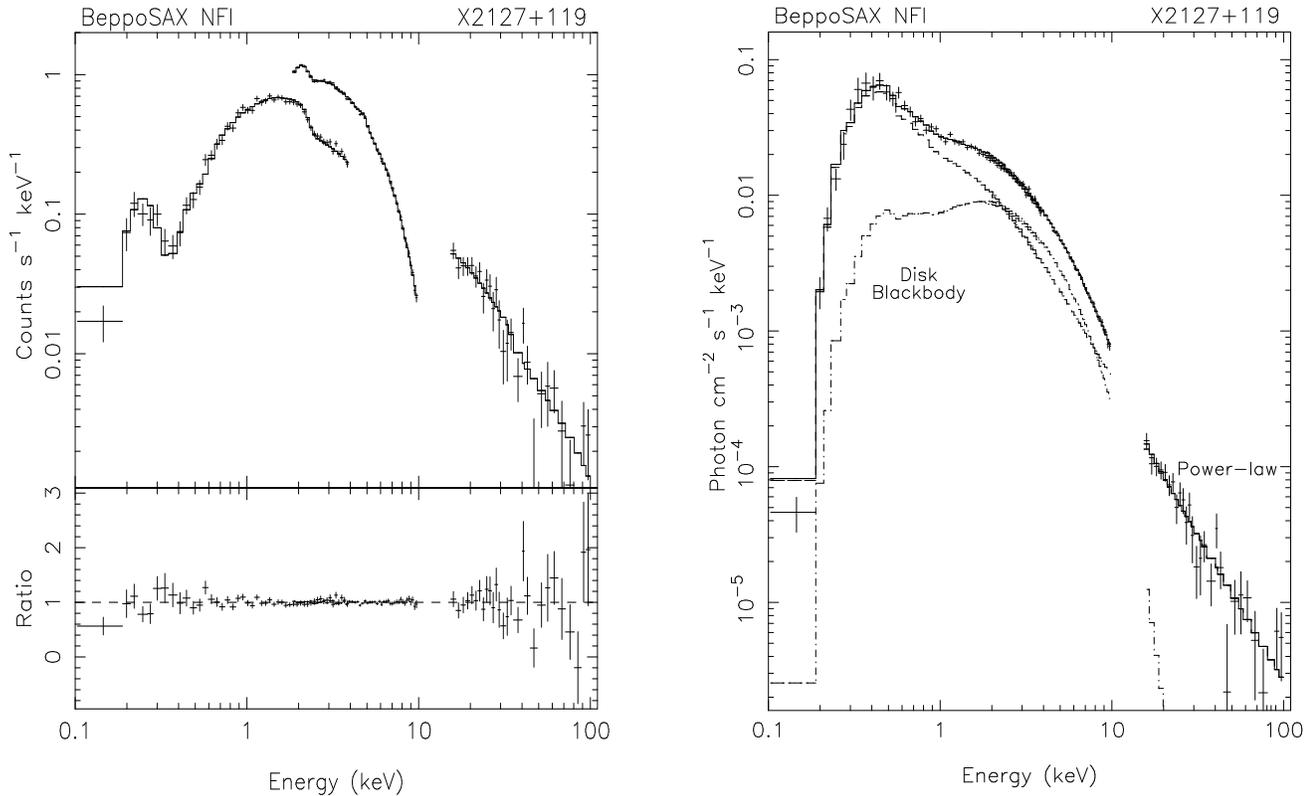

\mbox{ 
\hspace{-0.3cm}
\includegraphics[height=8.0cm,width=10.5cm,angle=-90]{h2114.f4}
\hspace{1.0cm}
\includegraphics[height=8.0cm,angle=-90]{h2114.f5}}
\vspace{0.5cm}
\caption[]{Best-fit partial covering model to the BeppoSAX broad band 
spectrum of \src. 
See Table~\ref{tab:spec_av} for the parameters}
\label{fig:ldarat}
\end{figure*}

A significantly better fit is obtained if a partial covering 
absorption
(the {\sc pcfabs} model in {\sc xspec}) replaces the different
absorbing column for the two components. 
This model consists 
of a column density fixed at the galactic value, together with
partial covering absorption for the previous two components
(a disk-blackbody and a power-law):

\[
{\rm Intensity = e^{-\sigma N_{gal}} \;[fe^{-\sigma N_H} \rm  +\; (1 -\; f)] \; ( I_{DBB} +\;  I_{PL}) ,}
\]
where \nh\  is the  intrinsic absorption for both  spectral
components and f is the covering fraction (${\rm 0 < f < 1}$).
The $\chi^2$ is significantly reduced to 147.5 for 125 dof.
The count and photon spectra are shown in Fig.~\ref{fig:ldarat} and the
best-fit parameters given in Table~\ref{tab:spec_av}.

In order to test the second possibility,
we replaced the partial covering absorption with
an ionized absorber ({\sc absori} in {\sc xspec}) in our best fit model.
The new model resulted in a worse fit, both when the iron
abundance was fixed at the \glob\ value ($\chi^2$ = 176.6 for 125 dof and
an ionization parameter L/nR$^{2}$ = $40 \pm 20$) and when
 fixed at the solar
value ($\chi^2$ = 166.3 for 125 dof and an ionization
parameter of $17\pm10$).  
 
The final possibility is the presence of an
additional soft component. To test this hypothesis, we included 
a bremsstrahlung (T $\sim 0.1$ keV), a blackbody, and a broad
Gaussian line to the disk-blackbody and power-law model.
None of these additional components significantly improved
the fit, giving values of $\chi^2$/dof
of 229/124, 890/127 and 245/123,
respectively. Thus, we exclude the possibility that the low-energy
residuals are caused by such an additional component.
We therefore conclude that of the models tried, the partial covering
gives the most reasonable fit and we refer to this subsequently as the 
best-fit model.
The lower limit to any high energy cutoff is
60~keV at 90\% confidence level. 
Since Fig.~\ref{fig:all} indicates that there are
strong energy dependent intensity variations,
we next investigated whether the best-fit model could be successfully
applied to individual intensity selected spectra.

\begin{table}
\begin{center}
\caption[]{Best-fit parameters for the overall BeppoSAX
spectrum. The model consists of a partial covering ({\sc pcfabs}) 
absorption, \nh, for a {\sc diskbb} and a power-law (see text).
Flux is in the 2--10~keV energy range and in units of
\ferg. The luminosity (2--10~keV) 
has been corrected for interstellar absorption and assumes a 
distance of   10.4~kpc (Durrel \& Harris \cite{dh:93})  }
\begin{tabular}{ll}
\hline
\noalign {\smallskip}
Parameter \\
\hline
\noalign {\smallskip}
${\rm N_{gal}}$ $(10^{20}$ atom cm$^{-2}$) &  $7 $ (fixed) \\ 
\nh\ $(10^{22}$ atom cm$^{-2}$)	&   $1.03 ^{+0.12} _{-0.09} $ \\ 
f                                   &   $0.64  ^{+0.05} _{-0.04} $  \\
kT$_{{\rm in}}$ (keV) 		&  $1.77 ^{+0.09} _{-0.07}$       \\
R$_{\rm in} (\cos \theta)^{0.5} / d_{10}$ (km)& $1.03 \pm 0.05$   \\
$\alpha$  		 	& $2.10 \pm 0.10  $    \\
I$_{{\rm PL}}$  		 &  $0.063 \pm 0.01 $      \\
Observed Flux                    &  $2.7\times10^{-10}$   \\
Luminosity (${\rm erg~s^{-1}})$   &   $3.3\times10^{36}$    \\
$\chi ^2$/dof                    &  147.5/125   \\
\noalign {\smallskip}                       
\hline
\label{tab:spec_av}
\end{tabular}
\end{center}
\end{table}

\subsubsection{Intensity selected spectra}
\label{sect:intspe}

\begin{figure}
  \centerline{\psfig{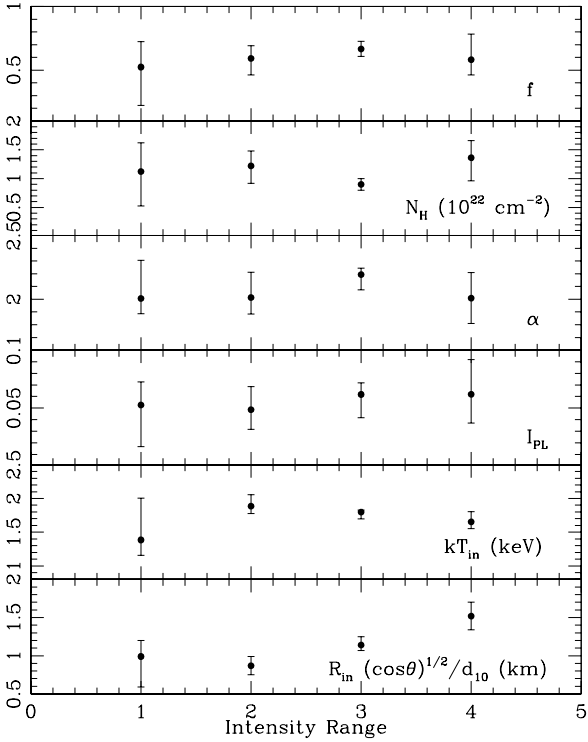}}
    \caption[]{Variation of the best-fit spectral parameters with 
intensity (see Table~\ref{tab:4spec}) }
\label{fig:par}
\end{figure}

A series of intensity selected spectra were produced.
Intervals corresponding to MECS 1.8--10~keV
count rates of $<$2.6, 2.6--3.6, 3.6--4.6, and $>$4.6~s$^{-1}$, 
when the data are accumulated with a binning of 256~s, were determined 
(see Fig.~\ref{fig:int}).
These intervals were used to extract a set of   four corresponding   LECS,
MECS and PDS spectra which
were rebinned and energy selected in the same way as in 
Sect.~\ref{subsect:overall}.
The different intensity ranges correspond approximately to
the deepest minimum (1), the broad secondary minimum (2), the secondary
maximum (3) and the brightest first maximum (4).
Since the power-law and disk-blackbody with a partial covering absorption
provides the best-fit to the overall
spectrum, this model was also fit to these new spectra.
Acceptable fits are also obtained in this case
and the results reported in Table~\ref{tab:4spec}
and Fig.~\ref{fig:par}.

\begin{figure}
  \centerline{\psfig{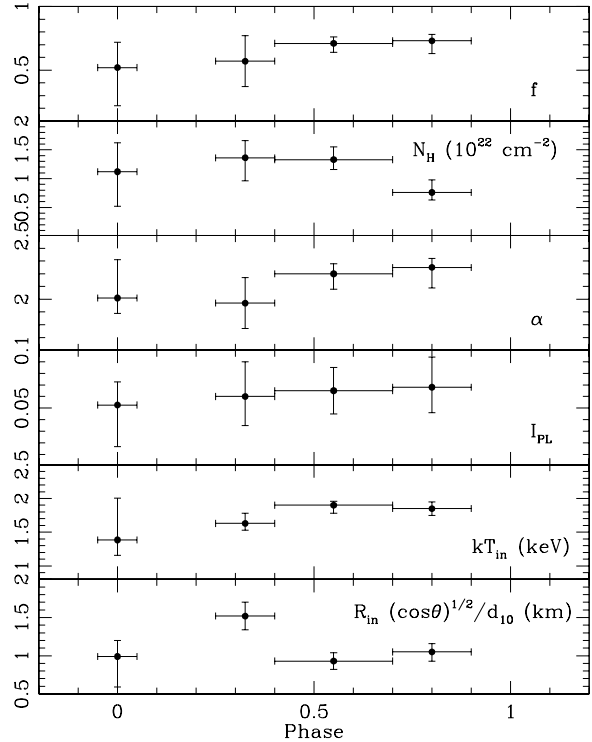}}
    \caption[]{Variation of the best-fit spectral parameters with 
orbital phase}
\label{fig:phase}
\end{figure}
 

Examination of the parameters given in Table~\ref{tab:4spec} shows
that ${\rm N_H}$, f, kT${\rm _{in}}$, $\alpha$, and I${\rm _{PL}}$
do not vary systematically with luminosity, but that the
normalization of the disk-blackbody 
(${\rm R_{in}}(\cos\theta)^{0.5}/{\rm d_{10}})$
increases significantly with increasing luminosity. 
See Fig.~\ref{fig:par}
for the dependence of the spectral parameters 
on intensity.
The disk-blackbody component contributes between 24\% and 62\% 
of the observed flux (uncorrected for the interstellar absorption)
in the 2--10 keV energy range. 
Thus, the energy dependent
variations shown in Figs.~\ref{fig:all} and \ref{fig:folded} may
be largely explained by variations in the normalization of this component.

\begin{table*}
\begin{center}
\caption[]{Best-fits to the BeppoSAX NFI spectra in intensity
intervals 1 to 4. 
The model consists of a partial covering ({\sc pcfabs}) absorption, 
\nh, for a {\sc diskbb} and a power-law (see text).
Fluxes are in the 2--10~keV energy range and in units of
$10^{-10}$~erg~cm$^{-2}$~s$^{-1}$. Luminosities (2--10~keV) 
are in units of $10^{36}$~erg~s$^{-1}$ and 
are corrected
for interstellar absorption and assume a distance of   10.4 kpc.
f${\rm _{bb}}$ is the fraction of the total observed flux 
contributed by the disk-blackbody component in the energy range 2--10 keV
 }
\begin{tabular}{lllll}
\hline
\noalign {\smallskip}
Parameter              & 1 & 2 & 3 & 4 \\
\hline
\noalign {\smallskip}
\nh\ ($ 10^{22}$ atom cm$^{-2}$)	  & $1.12 ^{+0.50} _{-0.60}$  &  $1.22 \pm 0.3$  &  $0.90 \pm 0.1 $ &   $1.36  ^{+0.3} _{-0.4} $  \\ 
f    & $0.52 ^{+0.2} _{-0.3}$  &  $0.59 ^{+0.10} _{-0.13}$  &  $0.67  \pm 0.06 $ &   $0.57  \pm 0.2 $ \\
kT$_{{\rm in}}$ (keV)   & $1.39 ^{+0.62} _{-0.23}$ &  $1.89 ^{+0.17} _{-0.11}$    &  $1.80 ^{+0.03} _{-0.1}$&   $1.63 ^{+0.15} _{-0.10}$     \\
R$_{\rm in} (\cos \theta)^{0.5} / d_{10}$ (km)& $0.99 ^{+0.21} _{-0.40}$  &  $0.87 \pm 0.12$   & $1.14 ^{+0.11} _{-0.07}$ &    $1.52 \pm 0.18$  \\
$\alpha$  		 	& $2.01^{+0.3} _{-0.12} $     & $2.01^{+0.2} _{-0.13} $  &  $2.19^{+0.05} _{-0.12}$   &  $1.97 \pm 0.2 $   \\
I$_{{\rm PL}}$  		& $0.053^{+0.020} _{-0.036} $  &$0.049 \pm 0.020 $  & $0.062 \pm 0.020 $ &  $0.060 \pm 0.030 $  \\
Observed Flux    &  $1.7$   & $2.4$   & $2.9$  & $3.5$    \\
Luminosity        &   $2.1$   & $3.0$    &  $3.6$  & $4.4$    \\
f${\rm _{bb}}$          &   $0.24$   & $0.53$    &  $0.62$  & $0.60$     \\
$\chi ^2$/dof    & 137.9/113     & 138/117  &  128/121 &  114.8/115  \\
\noalign {\smallskip}                       
\hline
\label{tab:4spec}
\end{tabular}
\end{center}
\end{table*}

 
\subsubsection{Phase selected spectra}
\label{sec:phase}

In order investigate the spectral variation as a function of
$\Phi$, four sets of phase selected spectra were extracted and fit with
the best-fit model. The results, 
shown in Fig.~\ref{fig:phase}, may be compared with 
the variations in spectral parameters with intensity shown in 
Fig.~\ref{fig:par}.
The four sets of spectra correspond to
primary minimum, primary maximum, 
secondary minimum and secondary maximum, respectively.
The column density does not display any obvious dependence on
$\Phi$, except possibly when the
source is in the secondary maximum.
As expected, the disk-blackbody normalization exhibits the most
obvious dependence on $\Phi$,  
showing a similar correlation to that with 
source intensity.


\subsection{ASCA spectrum}
\label{subsect:asca}

\begin{figure}
 
\centerline{\psfig{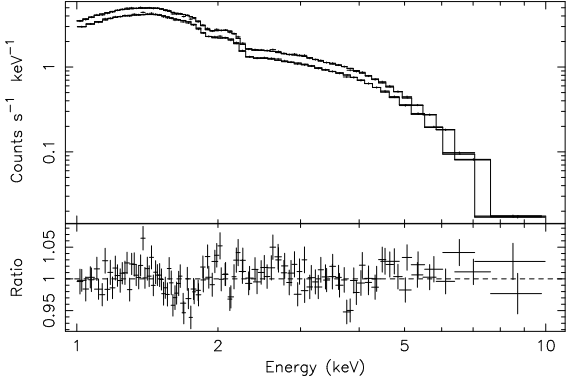}}
    \caption[]{Best-fit model to the ASCA-SIS spectrum of \src. 
               See Table~\ref{tab:asca} for the parameters}
   \label{fig:ldarat_asca}
\end{figure}

We next examined whether the BeppoSAX best-fit model presented above
is consistent with results from ASCA.
The best-fit BeppoSAX model was fit to the SIS 
spectra (see Fig.~\ref{fig:ldarat_asca}) and
the results are presented in Table~\ref{tab:asca}. 
We only considered data above 1 keV due to SIS calibration
uncertainties at lower energies (Hwang et al. \cite{h:99}).
Our results
show that the best-fit BeppoSAX model also provides a good fit
to the ASCA SIS spectra giving a $\chi ^2$ of 912.0 for 848 dof. 
This is a significantly better fit than when an
absorbed disk-blackbody plus a power-law model is used 
which gives a $\chi ^2$ of 1034 for 849 dof.
During the ASCA observation
the 2--10~keV flux   uncorrected for interstellar absorption was 
$3.0 \times 10^{-10}$~erg~cm$^{-2}$~s$^{-1}$,   within 10\%
of the value during the BeppoSAX observation,
and the disk-blackbody contributed 32\% of the total observed flux 
in this energy range.


\begin{table}
\begin{center}
\caption[]{Best-fit parameters for the overall ASCA
spectrum. The model consists of a partial covering 
({\sc pcfabs}) absorption, \nh, for a {\sc diskbb} 
and a power-law (see text).
Flux is in the 2--10~keV energy range and in units of
erg~cm$^{-2}$~s$^{-1}$. The luminosity (2--10~keV) 
has been corrected
for interstellar absorption and assumes a distance of   10.4  kpc}

\begin{tabular}{ll}
\hline
\noalign {\smallskip}
Parameter \\
\hline
\noalign {\smallskip}
${\rm N_{gal}}$ $(10^{20}$ atom cm$^{-2}$) &  $7 $ (fixed) \\ 
\nh\ $(10^{22}$ atom cm$^{-2}$)     &   $0.86  \pm 0.20$ \\ 
f                                   &   $0.76  \pm 0.03$  \\
kT$_{{\rm in}}$ (keV) 		    &  $1.96  ^{+0.20} _{-0.10}$       \\
R$_{\rm in} (\cos \theta)^{0.5} / d_{10}$ (km)& $0.70 \pm 0.12$   \\
$\alpha$  		    & $2.18 ^{+0.15} _{-0.21}  $    \\
I$_{{\rm PL}}$  		    &  $0.11 ^{+0.02} _{-0.03}$      \\
Observed Flux       &  $3.0\times10^{-10}$   \\
Luminosity          &   $3.7\times10^{36}$  \\
$\chi ^2$/dof   &  912/848   \\
\noalign {\smallskip}                       
\hline
\label{tab:asca}
\end{tabular}
\end{center}
\end{table}

\section{Discussion}
\label{sect:discussion}

We present results of a 1999 November BeppoSAX observation of \src\
covering two partial eclipses
and compare these with an earlier ASCA observation when the source
had a similar   2--10~keV intensity. 
The decrease in the 0.1--10~keV flux during the partial eclipses 
is smooth and gradual and lasts for at least 0.2 in phase 
and the secondary minimum  at $\Phi \sim 0.5$ 
is broader than the primary minima.
The overall variation in the hardness ratios HR1 and HR2
reveals a complex behaviour, with a large change of the HR1
during the partial eclipses. 
In particular, no spectral softening with increasing 
intensity is found,
contrary to the EXOSAT results of Callanan et al. (\cite{c:87}).

The 0.1--100~keV BeppoSAX spectrum is unusually complex and cannot be 
fit by any of the usual models applied to LMXRB 
such as an absorbed power-law and a blackbody. 
A good fit is obtained with the combination of a power-law 
and a disk-blackbody modified
by a partial covering absorption (with
additional absorption fixed at the interstellar value).
The same model provides a reasonable fit to an ASCA SIS spectrum
obtained in 1995. 
We have divided the BeppoSAX data into 4 intensity selected 
intervals and fit the above best-fit model to these data.
We find a positive correlation
between the disk-blackbody normalization and the source intensity with
the disk-blackbody contributing between 24\% and 62\% of the total 
2--10~keV flux. The absorption,
the blackbody kT,
$\alpha$, and the power-law normalization do not vary
systematically as the 2--10~keV source intensity varies
by a factor $\sim$2.

A phase selected spectral analysis also shows a correlation
of the disk-blackbody normalization with source intensity 
(see Fig.~\ref{fig:phase}, bottom panel).
The variability of the disk-blackbody component dominates 
the overall intensity and hardness variations of the source. 
The variation in  
hardness ratio HR1 during the partial eclipse 
is better modeled by a variation in the  
contribution of the disk-blackbody component. 
The absorbing column density  remains constant (within uncertainties) 
during and after the partial eclipse (Fig.~\ref{fig:phase}). 
 
The secondary minimum is one of the most variable and puzzling feature
of this source. Its location at $\Phi\sim0.5$   
is difficult to explain in terms of material located in a 
bulge formed by the impact of an accretion stream with the
outer edge of the the disk since this is
usually located at $\Phi\sim0.8$ (see e.g., White \& Holt \cite{wh:82};
Mason \& Cordova \cite{mc:82}).
I93 and Christian et al. (\cite{c:97}) suggest that the secondary
minimum is due to obscuring material located
in a ring at the circularization radius in 
the accretion disk, as in the model by Frank et al.
(\cite{fkl:87}).
However, we find no evidence for a higher ${\rm N_H}$
during the secondary minimum (see Fig.~\ref{fig:phase}). We caution
however, that the complex low-energy \src\ spectrum makes the
detailed interpretation of any spectral changes difficult. 

It is interesting
to speculate as to the physical nature of the different components
required to fit the \src\ spectrum.
The disk-blackbody component is most likely emission from the inner 
regions of the accretion disk and/or from a boundary layer 
(see e.g., Popham \&
Sunyaev \cite{ps:00}). This emission is
then scattered into the line of sight by the extended ADC. 
The power-law component is usually interpreted as being
due to the Comptonization of soft photons by hot electrons 
located in the ADC. 
In this case, the photon index translates into a Comptonization
y parameter of $\sim$0.7.
This non-thermal component does not show any evidence for a cut-off,
with a lower limit of 60~keV, indicating that the electron 
temperature must be $\approxgt$20~keV.

It is interesting to note that the ADC source X\thinspace1822--371
has also a very complex X-ray spectrum. If the BeppoSAX and ASCA
spectra are fit with simple power-law model then a strong deficit
is visible at $\sim$1.5~keV. This may be modeled as an edge at $\sim$1.3~keV
together with the intersection of power-law and blackbody components
(Parmar et al. \cite{p:00}). There is also low-energy complexity in the \src\
spectrum (see Fig.~\ref{fig:ldarat}), but no edge is required.
This difference may be related to the low abundance appropriate to \src.
It is suprising that the intensities of the two spectral components
in \src\ do not vary in the same way in the intensity selected fits.
This would be the case if the modulation was simply caused by the
geometric obscuration of a homogeneously emitting ADC. The fact that
the power-law normalization during the partial eclipses
is perfectly consistent (within the uncertainties) with
the values outside the partial eclipses (see Fig.~\ref{fig:par}), 
implies that the region emitting this component
is significantly larger than the companion star. 
On the other hand, we note that the uncertainties on the 
power-law normalizations
are large and an obscuration of an homogeneously emitting ADC
by a factor of $\sim$3 during the partial 
eclipse by the companion star (with the
estimated dimensions)  would be undetectable as a decrease of this same
factor 
in the power-law normalization (within the uncertainties).
 
The nature of the absorbed component required in both the
BeppoSAX and ASCA spectral fits is uncertain.
One possibility is that this is emission that passes through 
structure at the outer edge of the accretion disk after first being
scattered in the ADC. Rather than a single value of ${\rm N_H}$, the
measured valued would then represent an average, since a range
of absorbing columns would be expected in this case.

The orbital period P=17.11 hr implies the
presence of a (sub)giant companion 
(e.g., Verbunt \& Van den Heuvel \cite{vv:95}).
The parameters of the binary system can be derived from the orbital
period ${\rm P_{orb}}$ and assuming the mass ratio, q, between the mass 
of the companion
M${\rm _c}$ and the mass of the neutron star M${\rm _{ns}}$.
Taking M$_c = 0.8$ M$_\odot$ (the typical mass for a turn-off star
in a globular cluster) and M${\rm _{ns} = 1.4 M_\odot}$, the binary
separation is a = 3$\times10^{11}$~cm,   the radius R${\rm _c}$ 
of the companion star, filling the Roche lobe, is 
R${\rm _{c}}\sim10^{11}$~cm, the disk radius 
is R${\rm _{disk}\sim 10^{11}}$~cm, 
the diameter of the extended X-ray source (ADC) 
is $\sim1.5\times10^{11}$~cm (I93).

The nature of \src\ is intriguing. 
The fact that \src\ contains a neutron star
was firmly established by the detection 
of a radius expansion X-ray burst during a $Ginga$ observation in 1988 
(Dotani et al. \cite{d:90}, van Paradijs et al. \cite{v:90}).
Although the $Ginga$ Large Area Counter has a large field
of view (0.8${\rm ^o}$ x 1.7${\rm ^o}$), it is clear that
the burst came from \src\ since there was no other known X-ray
source in the field of view and  
between the precursor and the main
peak of the burst 
there was a significant reduction in the persistent X-ray flux.
Thus the burst must almost certainly have come from the source of the
persistent emission, \src.
The detection of such a luminous ($4.5 \times 10^{38}$~erg~s$^{-1}$ at
the peak, assuming isotropic emission)
burst indicates that at times 
the surface of the neutron star is directly observed.
The evidence that \src\ is primarily an ADC source comes mainly from
the extremely low ${\rm L_x/L_{opt}}$ ratio, 
typical of other ADC sources, the partial X-ray eclipses and
the optical spectrum which resembles that of other ADC sources
(e.g., Downes et al. 1996).

This apparantly discrepant nature
could be resolved if the highly energetic burst
ionized the absorbing material in the line of sight, 
as proposed by Smale et al. (1992) for an X-ray burst
observed from the dipping LMXRB X\thinspace1916--053.
Following Smale et al. (1992), the number of photons per unit
area emitted by the
burst and available to ionize the obscuring material can be calculated
from ${\rm N_{ph} = L\Delta t / \epsilon 4\pi R^{2}}$, where L is the
luminosity of the burst, $\Delta$t the duration of the most luminous
part of the burst, $\epsilon$ is the mean X-ray energy, and R the distance
of the obscuring material to the source.
Taking ${\rm L \sim 10^{38}}$ ergs~s$^{-1}$, $\Delta$t $\sim$ 1~s, 
$\epsilon \sim $1~keV, and
R$ \sim 10^{11}$~cm, gives 
N${\rm _{ph} \sim 5\times 10^{23}}$~photons~cm$^{-2}$,
a hundred times larger than the intrinsic column density determined
here.
So, material located in the outer regions of the accretion disk could have
been temporarily ionized, allowing the neutron star to be viewed directly.

\begin{acknowledgements}
The \sax\ satellite is a joint Italian-Dutch programme. 
L. Sidoli acknowledges an ESA Fellowship. 
This research has made use of data obtained through the High Energy 
Astrophysics Science Archive Research Center Online Service, provided 
by the NASA/Goddard Space Flight Center.
\end{acknowledgements}


\begin{thebibliography}{}

\bibitem[1989]{a:89}
Anders E., Grevesse N., 1989, Geochimica et Cosmochimica Acta 53, 197 

\bibitem[1984]{a:84}
Auri\`ere M., le F\`evre O., Terzan A., 1984, A\&A 138, 415
  
\bibitem[1997]{b:97}
Boella G., Chiappetti L., Conti G., et al., 1997, A\&AS 122, 327

\bibitem[1987]{c:87}
Callanan P.J., Fabian A.C., Tennant A.F., Redfern R.M., Shafer R.A.,
1987, MNRAS 224, 781

\bibitem[1999]{c:99}
Callanan P.J., Drake J.J., Fruscione A., 1999, ApJ 521, 125

\bibitem[1986]{c:86}
Charles P.A., Jones D.C., Naylor T., 1986, Nat 323, 417

\bibitem[1997]{c:97}
Christian D.J., Smale A.P., Swank J.H., Serlmitsos P.J., 1997,
ApJ 477, 424
 
\bibitem[1990]{d:90}
Dotani T., Inoue H., Murakami T., et al., 1990, Nat 347, 534

\bibitem[1996]{d:96}
Downes R.A., Anderson S.F., Margon B., 1996, PASP 108, 688

\bibitem[1993]{dh:93}
Durrell P.R., Harris W.E., 1993, AJ 105, 1420

\bibitem[1987]{fkl:87}
Frank J., King A.R., Lasota J.P., 1987, A\&A 178, 137

\bibitem[1997]{f:97}
Frontera F., Costa E., Dal Fiume D., et al., 1997, A\&AS 122, 371

\bibitem[1992]{g:92}
Geisler D., Minniti D., Claria J., 1992, AJ 104, 627

\bibitem[1998]{g:98}
Guainazzi M., Parmar A.N., Segreto A., Stella L., Oosterbroek T.,
1998, A\&A 339, 802

\bibitem[1999]{g:99}
Guainazzi M., Parmar A.N., Oosterbroek T., 1999, A\&A 349, 819

\bibitem[2000]{g:00}
Guainazzi M., Parmar A.N., Oosterbroek T., 2000, Ap. Lett \& Comm. submitted

\bibitem[1983]{hg:83}
Hertz P., Grindlay J.E., 1983, ApJ 275, 105

\bibitem[1998]{h:98}
Homer L., Charles P.A., 1998, New Astron. 7, 435 (HC98)

\bibitem[1999]{h:99}
Hwang U., Mushotzky R.F., Burns J.O., et al., 1999, ApJ 516, 604

\bibitem[1993]{i:93}
Ilovaisky S.A., Chevalier C., Auri\`ere M., et al., 1993, A\&A 270, 139 (I93)

\bibitem[1986]{m:86}
Makishima K., Maejima Y., Mitsuda K., et al., 1986, ApJ 285, 712

\bibitem[1982]{mc:82}
Mason K.O., Cordova F.A., 1982, ApJ 262, 253

\bibitem[2000]{mfr:99}
Merloni A., Fabian A.C., Ross R.R., 2000, MNRAS 313, 193

\bibitem[1984]{m:84}
Mitsuda K., Inoue H., Koyama K., et al., 1984, PASJ 36, 741

\bibitem[1983]{m:83}
Morrison D., McCammon D., 1983, ApJ 270, 119
 

\bibitem[1997]{p:97} 
Parmar A.N., Martin D.D.E., Bavdaz M., et al., 1997, A\&AS 122, 309

\bibitem[1999]{p:99} 
Parmar A.N., Oosterbroek T., Guainazzi M., et al., 1999, A\&A 351, 225

\bibitem[2000]{p:00} 
Parmar A.N., Oosterbroek T., Del Sordo S., et al., 2000, A\&A 356, 175 
 
\bibitem[2000]{ps:00}
Popham R., Sunyaev R., 2000, preprint (astro-ph/0004017), ApJ submitted

\bibitem[1995]{ps:95}
Predehl P., Schmitt J.H.M.M., 1995, A\&A 293, 889

\bibitem[1992]{s:92}
Smale A.P., Mukai K., Williams O.R., Jones M.H.,
Corbet R.H.D., 1992, ApJ 400, 330

\bibitem[1991]{s:91}
Sneden C. Kraft R.P., Prosser C.F., Langer G.E., 1991, AJ 102, 2001

\bibitem[1994]{t:94}
Tanaka Y., Inoue H., Holt S.S., 1994, PASP 46, L37
 
\bibitem[1990]{v:90}
Van Paradijs J., Dotani T., Tanaka Y., Tsuru T., 1990, PASJ 42, 633

\bibitem[1995]{vv:95}
Verbunt F., van den Heuvel E.P.J., 1995, in ``X--ray Binaries", 
Lewin W.H.G., van Paradijs J.,
van den Heuvel E.P.J. eds., Cambridge Univ. Press,
Cambridge, p. 457

\bibitem[1982]{wh:82}
White N.E., Holt S.S., 1982, ApJ 257, 318
 


\end{thebibliography}
\end{document}